\documentclass[usenatbib]{mnras}
\usepackage{graphics,graphicx}
\usepackage{amsmath}
\usepackage{amssymb,latexsym,mathrsfs, bm}
\usepackage{color}
\usepackage{float}
\usepackage{natbib}
\usepackage{times}
\usepackage{pdflscape}
\usepackage{longtable}

\def\eq#1{{Eq.~(\ref{#1})}}

\title[Cross-correlation forecasts]{Cross-correlating 21 cm and galaxy surveys: 
implications for cosmology and astrophysics}
\author[Padmanabhan, Refregier and Amara]{Hamsa
Padmanabhan$^{1,2}$\thanks{Electronic address: hamsa@cita.utoronto.ca},
Alexandre Refregier$^2$\thanks{Electronic address:
{alexandre.refregier@phys.ethz.ch}},
Adam Amara$^{2,3}$\thanks{Electronic address: {adam.amara@port.ac.uk}}\\
$^{1}$ Canadian Institute for Theoretical Astrophysics, 60 St. George St., 
Toronto, ON, M5S 3H8, Canada \\
$^{2}$ Institute for Particle Physics and Astrophysics, ETH Zurich, 
Wolfgang-Pauli-Strasse 27, CH-8093 Z\"{u}rich, Switzerland \\
$^{3}$ Institute of Cosmology \& Gravitation, University of Portsmouth, Dennis Sciama Building, Burnaby Road, Portsmouth PO1 3FX, UK}

\begin{document}
\date{ }
\maketitle

\begin{abstract}
We forecast astrophysical and cosmological parameter constraints from synergies 
between 21 cm intensity mapping and wide field optical galaxy surveys (both 
spectroscopic and photometric) over $z \sim 0-3$. We focus on the following 
survey combinations in this work:  (i) a CHIME-like and DESI-like survey in 
the northern hemisphere, (ii) an LSST-like and SKA I MID-like survey and (ii) a 
MeerKAT-like and DES-like survey in the southern hemisphere. We work with the 
$\Lambda$CDM cosmological model having  parameters $\{h, \Omega_m, n_s, 
\Omega_b, \sigma_8\}$,  parameters $v_{c,0}$ and $\beta$ representing the 
cutoff 
and slope of the HI-halo mass relation in the previously developed HI halo 
model 
framework, and a parameter  $Q$ that represents the scale dependence of the 
optical galaxy bias. Using a Fisher forecasting framework, we explore (i) the 
effects of the HI and galaxy astrophysical uncertainties on the cosmological 
parameter constraints, assuming priors from the present knowledge of the 
astrophysics, (ii)
the improvements on astrophysical constraints over their current priors in the 
three configurations considered, (ii) the tightening of the constraints on the 
parameters relative to the corresponding HI auto-correlation surveys alone.

\end{abstract}


\section{Introduction}

Intensity mapping of redshifted emission lines \citep{bharadwaj2001, loeb2008} 
is a novel technique that has 
the potential to perform precision cosmology by detecting the integrated 
emission from sources across redshifts without resolving individual systems 
\citep[e.g.,][]{kovetz2019}. Besides offering rich insights into the physics of 
star formation history and the processes governing galaxy evolution 
\citep[e.g.,][]{wyithe2008, wolz2016}, it has the 
ability to improve vastly upon the current measurements of cosmological 
parameters \citep[e.g.,][]{bull2014}, as well as place competitive constraints 
on inflationary scenarios and physics beyond the standard model 
\citep[e.g.,][]{hall2013, camera2013, pourtsidou2016, masui2010}. The most well-studied 
example of line-intensity mapping involves that of the redshifted 21-cm 
emission of neutral hydrogen (hereafter, HI) which arises primarily in 
star-forming galaxies and the intergalactic medium at low to moderate 
redshifts. 

Using line-intensity mapping techniques in synergy with other, more traditional 
and established tracers of large-scale structure is crucial to unlock the true 
potential of these surveys. It is known that  \citep[e.g.,][]{seljak2009, fonseca2015, fonseca2017, camera2015}  
cross-correlations of several individual tracers of the cosmological structure 
often offer several significant advantages over individual surveys. { The 
systematic survey-specific effects are mitigated to a large extent, the noise 
in the surveys is reduced, and the foregrounds and contaminants of individual 
surveys are, 
in most cases, uncorrelated and hence do not bias the cross-correlation 
measurement. \footnote{ However, we note that the presence of foregrounds may greatly increase the variance, which is an important effect, and hence efficient foreground cleaning or avoidance techniques are still necessary in order to isolate the true signal.}} The cosmic variance can be mitigated in the measurement of 
some of the cosmological parameters \citep[e.g.,][]{mcdonald2009, abramo2013}. 

To this end, cross-correlating 21 cm intensity mapping surveys with optical 
galaxies offers rich possibilities into exploiting the complementarity of both 
approaches. The first intensity mapping detection of the redshifted 21 cm 
emission at $z \sim 0.53-1.12$ with the Green Bank Telescope (GBT) was made in 
cross-correlation with the DEEP2 optical galaxy survey \citep{chang2013}, and 
has been followed up since then resulting an updated cross-power spectrum using 
the WiggleZ survey at $z \sim 0.8$ \citep{masui13} and an upper limit on the 
auto-power spectrum \citep{switzer13}, which was used to place the first 
intensity mapping constraints on the product of the neutral hydrogen density 
and bias parameter. Similarly, the cross-power spectrum between 2dF galaxies in 
the southern hemisphere and the Parkes HI intensity field at $z \sim 0$ has 
also been measured recently \citep{anderson2018}, offering insights into the 
clustering of low redshift HI systems. 

The auto-correlation power spectrum signal of high-redshift 21 cm in emission has yet to 
be observed, although there are several experiments planned or in the final 
stages of commissioning to achieve this goal. These include (i) the Canadian 
Hydrogen Intensity Mapping Experiment 
(CHIME)\footnote{https://chime-experiment.ca/}, (ii) the Hydrogen Intensity and 
Real-time Analysis eXperiment 
(HIRAX)\footnote{https://www.acru.ukzn.ac.za/~hirax/}, (iii) the BAO In Neutral Gas 
Observations \citep[BINGO;][]{battye2012}, (iv) the TianLai experiment 
\citep{chen2012} and the Five hundred metre Aperture Spherical Telescope 
\citep[FAST;][]{smoot2017},
(v) the Meer-Karoo Array Telescope \citep[MeerKAT;][]{jonas2009} and
(vi) the Square Kilometre Array (SKA) Phase I MID 
\footnote{http://www.ska.ac.za/}.

The optical surveys of key interest for cross-correlations with the 21 cm 
surveys planned above include those with (i) the completed Dark Energy Survey 
(DES)\footnote{https://www.darkenergysurvey.org/}, a photometric 
galaxy survey over $z\sim 0.5 -1.4$,  cataloging hundreds of millions of 
galaxies in the southern hemisphere,  (ii) the forthcoming Dark Energy 
Spectroscopic Instrument (DESI), a spectroscopic survey which will target a few 
tens of millions of galaxies in the northern sky over the redshift range $z 
\sim 
0-3$, measuring cosmological parameters and the growth of structure through 
redshift space distortions, and (iii) future galaxy surveys conducted with the 
Large Synoptic Survey Telescope (LSST)\footnote{www.lsst.org}, and the 
space-based Euclid\footnote{www.euclid-ec.org} spectroscopic survey.

In \citet{pourtsidou2016}, synergies between a MeerKAT HI intensity mapping 
survey and photometric galaxies from the  Dark Energy Survey (DES) have been 
explored. Recent studies \citep{mona2019, amadeus2019} have illustrated the 
ability of HI intensity mapping (with SKA and HIRAX) in cross-correlation with 
photometric galaxy surveys (such as DES and LSST) to measure the gravitational 
lensing magnification. In \citet{carucci2017}, the synergies between 21 cm SKA 
I MID  and the Baryon Oscillation Spectroscopic Survey (BOSS)-like Lyman-alpha 
surveys have been presented, which can constrain the bias of astrophysical 
systems. In \citet{cosmicvisions2018}, various prospects for cross-correlating 
21 cm intensity mapping and optical surveys have been explored, including with 
QSOs observed by the DESI survey.  It has been shown \citep{chen2018} that a 
combining a CMB Stage 4-like survey with 21 cm intensity mapping observations 
from a SKA I MID like survey cross-correlated with DESI quasars can enable 
precise measurements of the growth factor, and test the predictions of general 
relativity on the largest scales.
\citet{witzemann2018} explore how synergies between a SKA I MID like survey and 
a photometric LSST survey can mitigate the effects of cosmic variance, enabling 
measurements of the bias ratio at large scales up to $\ell \sim 3$, and 
\citet{hall2017} illustrate how peculiar velocity effects can be constrained 
using the dipole of the redshift space cross-correlation between 21 cm and 
optical surveys conducted with various experiments. In \citet{ballardini2019a}, the constraints for local primordial non-Gaussianity
has been studied with a SKA-like intensity mapping survey in
cross-correlation with photometric galaxy surveys (Euclid and LSST)
and CMB lensing.

In this paper, we build upon our previous forecasting analyses in 
\citet[][hereafter Paper I]{hparaa2018} which focussed on auto-correlation 21 
cm power spectra and extend these to the case of measurement of the 
astrophysical and cosmological parameters using a combination of 21 cm and 
optical galaxy surveys. We use the uncertainties in the parameters coming from 
the combination of current measurements to set realistic priors on the HI 
astrophysics. We consider three sets of surveys in this work: (i) a CHIME-like 
survey overlapping with DESI in the northern hemisphere, (ii) a MeerKAT-like 
survey overlapping with DES in the southern hemisphere and (iii) a SKAI MID 
survey overlapping with the Large Synoptic Survey Telescope (LSST), again in 
the southern hemisphere.

The paper is organized as follows. For modelling the HI distribution and 
density profile, we use the halo model framework introduced in \citet{hpar2017} 
and expanded upon in \citet[][hereafter Paper II]{hparaa2016}, which describes 
the best fitting HI-halo mass relation and profile  constrained by the 
currently available data. The bias and redshift distribution of the optical 
galaxies are modelled 
following 
the treatment for the particular survey under consideration. These frameworks 
are 
briefly described in  Sec. \ref{sec:framework}. Using the cross-correlation 
power spectrum thus derived, we compute the relative errors on the 
astrophysical and cosmological parameters under consideration using a Fisher 
forecasting formalism for the three survey sets in Sec. \ref{sec:surveys}.  We 
comment on the comparison of these predictions to those from the corresponding 
21 cm auto-correlation 
constraints, and summarize our conclusions in Sec. \ref{sec:conclusions}.

\section{Framework for Fisher forecasts}
\label{sec:framework}
We use the halo model for neutral hydrogen (see Paper II) and build upon our 
existing forecasts in Paper I, which had focused on the HI auto-correlation 
surveys alone. The halo model framework consists of a prescription assigning 
average HI mass to halo mass $M$ at redshift $z$, given by:
\begin{eqnarray}
M_{\rm HI} (M,z) &=& \alpha f_{\rm H,c} M \left(\frac{M}{10^{11} h^{-1} 
M_{\odot}}\right)^{\beta} \nonumber \\
&\times& \exp\left[-\left(\frac{v_{c0}}{v_c(M,z)}\right)^3\right] 
\end{eqnarray}
In the above formula, the free parameters are given by: (i) $\alpha$, the 
average HI fraction relative to cosmic $f_{\rm H,c}$, (ii)   $\beta$, the logarithmic slope which represents the deviation from linearity of the 
HI-halo mass prescription, and (iii) $v_{c0}$, which denotes the minimum virial 
velocity below which haloes preferentially do not host HI. 

To model the smaller scales in the HI power spectrum, we also need a 
prescription for the profile of the HI as a function of radius, halo mass and 
redshift, which is found to be well modelled by an exponential function { (Paper II, see also the observational results from, e.g., \citet{bigiel2012})}:
\begin{equation}
\rho(r,M) = \rho_0 \exp(-r/r_s)
\label{rhodefexp}
\end{equation}
with the scale radius $r_s$ given by:
\begin{equation}
 r_s = R_v(M,z)/c_{\rm HI} (M,z)
 \end{equation}
with $R_v$ being the halo virial radius and $c_{\rm HI}$ being the 
concentration parameter of the HI systems, which is analogous to the 
corresponding expression for dark matter:
\begin{equation}
 c_{\rm HI}(M,z) =  c_{\rm HI, 0} \left(\frac{M}{10^{11} M_{\odot}} 
\right)^{-0.109} \frac{4}{(1+z)^{\gamma}}.
\end{equation}
The HI profile thus introduces two more free parameters through the 
concentration parameter and its evolution: (i) $c_{\rm HI, 0}$ representing the 
overall normalization and $\gamma$ which encodes the evolution of the function 
with redshift.
In order to compute the nonlinear HI power spectrum, we need the Fourier 
transform of the profile function, given by:
\begin{equation}
u_{\rm HI}(k|M) = \frac{4 \pi}{M_{\rm HI} (M)} \int_0^{R_v} \rho_{\rm HI}(r) 
\frac{\sin kr}{kr} r^2 \ dr
\end{equation}
which allows us to write the power spectrum for HI intensity fluctuations as 
the sum of the 1- and 2-halo terms:
\begin{equation}
P_{\rm HI} (k,z) = P_{\rm 1h, HI} + P_{\rm 2h, HI}
\end{equation}
with
\begin{equation}
P_{\rm 1h, HI} (k,z) = \frac{1}{\bar{\rho}_{\rm HI}^2} \int dM \  n(M) \ M_{\rm 
HI}^2 \ |u_{\rm HI} (k|M)|^2
\label{power1h}
\end{equation}
and

\begin{eqnarray}
&& P_{\rm 2h, HI}  (k,z) =  P_{\rm lin} (k) \nonumber \\
&&\left[\frac{1}{\bar{\rho}_{\rm HI}} \int dM \  n(M) \ M_{\rm HI} (M) \ b (M) 
\ |u_{\rm HI} (k|M)| \right]^2
\label{power2h}
\end{eqnarray}
with $P_{\rm lin}(k)$ being the linear matter power spectrum.

For  computing the power spectrum of the optical galaxies in the survey, we use 
the expression:

\begin{equation}
P_{\rm gal} (k,z) = P_{\rm dm} (k,z) \ b^2_{\rm gal} (k,z)
\end{equation}

where $P_{\rm dm}$ is the dark matter power and the $b^2_{\rm gal} (k,z)$ denotes 
the galaxy-galaxy bias factor. This factor changes according to the survey and 
the type of galaxies under consideration.
This (scale-dependent bias) is modelled following the parameters given by 
\citet{amendola2015} which is based on the $Q$-formula of \citet{cole2005}:
\begin{equation}
b_{\rm gal} = b_{\rm gal, ls} b_0 \left(\frac{1 + Q k^2}{1 + Ak} \right)^{1/2}
\end{equation}
where the values $b_0 = 1.3, A = 1.7, Q = 4.6$ are assumed not to vary with 
redshift. The $b_{\rm gal, ls}$ term depends on the survey under consideration. 

The angular power spectrum on the sky for HI is computed by using the standard 
result:
\begin{eqnarray}
			 &&C_{\ell,\rm{HI}}(z,z') = \frac {2} {\pi} \int 
d\tilde z\, W_{\rm HI}(\tilde z) \int d\tilde z'\, W'_{\rm HI}(\tilde z')  
\nonumber \\
			&& \hskip-0.2in \times  \int k^2 dk\, \langle \delta_{\rm HI} 
(\textbf{k}, z) \delta_{\rm HI} (\textbf{k}', z') \rangle 
j_\ell(kR(\tilde z)) j_\ell(kR(\tilde z')),
	\label{eq:lintheory}
\end{eqnarray}
In the above expression, $\langle \delta_{\rm HI} (\textbf{k}, z) \delta_{\rm 
HI} (\textbf{k}', z') \rangle$ is the ensemble average of the HI density 
fluctuations at $(\textbf{k},z)$ and $(\textbf{k}', z')$ respectively. This is, 
in general, not expressible purely in terms of the power spectrum of HI as 
defined above, $P_{\rm HI} (k,z)$ evaluated at either of $\{z,z'\}$ since the 
density field evolves with $z$. However, in many cases, 
one can approximate this as $P_{\rm HI} (k,z_{m})$ where $z_m$ is the mean 
redshift of the given bin. In what follows, we use $z$ and $z_m$ 
interchangeably.

In the above expression, the $W_{\rm HI}, W'_{\rm HI}$ are the HI window 
functions at the redshifts 
$z$ and $z'$, taken 
to be uniform across the redshift bin considered, and $R(z)$ is the co-moving 
distance to redshift $z$. We use a 
top hat window function $W_{\rm HI}(z)$ with a width of $\Delta z  = 0.5$.

For a generic galaxy survey, calculation of the angular power spectrum yields 
the expression:
\begin{eqnarray}
			 &&C_{\ell,\rm{gal}}(z,z') = \frac {2} {\pi} \int 
d\tilde z\, W_{\rm g}(\tilde z) \int d\tilde z'\, W'_{\rm g}(\tilde z')  
\nonumber \\
			& &\times  \int k^2 dk\, P_{\rm gal}(k,z) 
j_\ell(kR(\tilde z)) j_\ell(kR(\tilde z')),
	\label{eq:lintheory}
\end{eqnarray}
The dark matter power spectrum for linear scales can alternatively be written 
as $P_{\rm dm} (k,z) 
= P_{\rm dm} (k,0) D^2(z)$ where
$D(z)$ is the growth factor for the dark matter 
perturbations whose power spectrum is normalized such that $D(0) = 1$. 
The window function for the galaxy survey, $W_{\rm g}$, can be different 
from that of the HI, and depends on the details of the selection function 
(usually denoted by $\phi(z)$) of each survey. 
 Parametrized forms for $\phi(z)$ are available for different galaxy surveys 
and usually follow a standard functional form \citep{smail1995}:
\begin{equation}
 \phi(z) \propto z^{\alpha} \exp(- (z/z_0)^{\beta})
 \label{selfunc}
\end{equation} 
where $\alpha$, $\beta$ and $z_0$ are fitted from the galaxy counts data in 
different redshift bins.
Once $\phi(z)$ is known, we derive the window function for the survey as:
\begin{equation}
 W_{\rm g}(z) = \phi(z)/\int_{z_{\rm min}}^{z_{\rm max}} \phi(z) dz
\end{equation} 
where $z_{\rm min}$ and $z_{\rm max}$ are the redshift edges of the survey. 
This ensures that the window function is normalized, i.e. 
\begin{equation}
 \int_{z_{\rm min}}^{z_{\rm max}} W_g(z) dz = 1
\end{equation}

 The calculation of the angular power spectra above, both for HI and for 
galaxies, can be simplified on using the 
Limber approximation \citep{limber1953} which is a good approximation in the 
large $\ell$ ($\ell > 50$) limit. The expression can be shown to reduce to:
\begin{equation}
C_{\ell, {\rm HI/gal}} = \frac{1}{c} \int dz  \frac{{W_{\rm HI/gal}}(z)^2 
H(z)}{R(z)^2} 
P_{\rm HI,gal} [\ell/R(z), z]
\label{cllimber}
\end{equation}.

The cross-correlation signal is then calculated as:
\begin{equation}
C_{\ell, \times} = \frac{1}{c} \int dz \frac{{W_{\rm HI}(z) W_{\rm gal}(z)} 
H(z)}{R(z)^2} 
(P_{\rm HI} P_{\rm gal})^{1/2}
\label{clcross}
\end{equation}
where the arguments of both power spectra ($P_{\rm HI}$ and $P_{\rm gal}$) are 
at $[\ell/R(z), z]$.
{ Noise in the HI intensity mapping survey, for the MeerKAT-like and SKA I MID-like configurations, is calculated using the standard 
expression assuming the interferometer array to operate in the single-dish 
autocorrelation mode \citep[e.g.,][]{Knox1995,ballardini2019}:
\begin{equation}
N_{\ell, \rm HI} = \left(\frac{T_{\rm sys}}{\bar{T}(z)}\right)^2 
\left(\frac{\lambda_{\rm obs}}{D_{\rm dish}}\right)^2 \left(\frac{1}{2 N_{\rm 
dish} t_{\rm pix} \Delta \nu}\right) W_{\rm beam}^2(\ell)
\label{noise}
\end{equation}
In the above expression, $N_{\rm dish}$ denotes the number of interferometer 
dishes, each assumed to have the diameter $D_{\rm dish}$, and $\lambda_{\rm 
obs}$ is the observed wavelength. The  $\bar {T}(z)$ is the mean 
brightness temperature at redshift $z$ defined by:
\begin{equation}
	\bar T(z) \simeq {44} \ {\mu {\rm K}} \left(\frac{\Omega_{\rm 
HI}(z)h}{2.45\times10^{-4}} \right)\frac{(1+z)^2}{E(z)}
	\label{eq:tbar}
\end{equation}
where $E(z)=H(z)/H_0$ is the normalized Hubble parameter at that redshift. 
The $T_{\rm sys}$ is the system temperature, calculated following $T_{\rm 
sys} = T_{\rm inst} \ +  \ 60 \ {\rm K} \left(\nu/300  \ {\rm MHz} 
\right)^{-2.5}$ where $T_{\rm inst}$ is the instrument temperature and $\nu$ is 
the observing frequency.
The quantity $W_{\rm beam}^2(\ell)$ denotes the beam window function due to the finite angular resolution of the instrument operating in single-dish mode \citep[e.g.,][]{pourtsidou2016, pourtsidou2016a}, and is given by:
\begin{equation}
 W_{\rm beam}^2(\ell) = \exp\left[\frac{\ell (\ell + 1) \theta_B^2}{8 \ln 2}\right]
\end{equation}
where $\theta_B \approx \lambda_{\rm obs}/D_{\rm dish}$ is the beam FWHM of a single dish.

The integration time per beam is $t_{\rm pix}$ and the $\Delta \nu$ denotes 
the frequency band channel width, which is connected to the tomographic 
redshift 
bin separation $\Delta z$. For the purposes of the noise calculation, we assume that
$\Omega_{\rm HI}(z)h = 2.45\times10^{-4}$, independent of redshift.

For the CHIME-like experiment we consider, we use the full interferometric noise expression, which is given by:
\begin{equation}
 N_{\ell, {\rm CHIME}}  = \frac{4 \pi f _{\rm sky}}{{\rm FoV}n_{\rm base}(u)n_{\rm pol} N_{\rm beam}  t_{\rm tot} \Delta \nu} \left(\frac{T_{\rm sys}}{\bar{T}}\right)^2 
\left(\frac{\lambda_{\rm obs}^2}{A_{\rm eff}}\right)^2.
\label{noisechime}
\end{equation}

In the above equation, $N_{\rm beam}$ is the number of independent beams, and for CHIME, $N_{\rm beam} = N_f \times N_{\rm cyl}$ where $N_f = 256$ is the number of feeds, and $N_{\rm cyl} = 4$ is the number of cylinders.  Each is assumed to have the effective area $A_{\rm eff} = \eta L_{\rm cyl} W_{\rm cyl}/N_f$, where $\eta = 0.7$ , $W_{\rm cyl} = 20\,\mathrm{m}$ is the width of each cylinder, and $L_{\rm cyl} = 100\,\mathrm{m}$ is its length.  The total integration time, denoted by $t_{\rm tot}$ is taken to be 1 year for the CHIME-like survey considered here. The $n_{\rm pol}$ is the number of polarisation channels (taken to be 2). The baseline number density is $n_{\rm base}(u)$, expressible in terms of the multipole $\ell$ via $u=\ell/(2\pi)$. This quantity is approximated as independent of $u$ up to a maximum baseline length $u_{\rm max}$, viz. $n_{\rm base}(u) = N_{\rm beam}^2/(2 \pi u_{\rm max}^2)$. The $u_{\rm max}$ denotes the longest baseline $d_{\rm max}$ measured in wavelength units, $u_{\rm max} = d_{\rm max}/\lambda_{\rm obs}$,  with $d_{\rm max} = 269\,\mathrm{m}$ for CHIME \citep{obuljen2018}. The field of view for the CHIME interferometer is approximated as ${\rm FoV} \approx \pi/2 \times \lambda/W_{\rm cyl}$ \citep{newburgh2014} for this configuration. 

}

The noise for the galaxy survey is taken to be the (Poisson) shot noise, 
calculated as $N_{\ell, {\rm gal}} = n_{\rm gal, bin}^{-1}(z)$ where $n_{\rm 
gal, bin}(z)$ is 
the number density of galaxies per steradian in the bin centred at redshift 
$z$.  Given the selection function of the galaxies, $\phi(z)$ defined in 
\eq{selfunc}, this quantity is computed as:
\begin{equation}
 n_{\rm gal, bin} (z) = \int_{z - \Delta z /2}^{z + \Delta z /2} \phi(z') dz'
\end{equation}

Finally, the variance of the forecasted angular power spectrum is calculated as:
\begin{eqnarray}
&& (\Delta C_{\ell, \times})^2 = \frac{1}{(2 \ell + 1)  \Delta \ell f_{\rm sky, 
\times}}  \nonumber \\
 &&\left[\left(C_{\ell, \rm HI} + N_{\ell, \rm HI} \right) \left(C_{\ell, \rm 
gal} + N_{\ell, \rm gal} \right) 
+  C_{\ell, \times}^2  \right]
\label{clerror}
\end{eqnarray}

In the above expression, the quantity $f_{\rm sky, \times}$ denotes the sky 
coverage of the overlap between the surveys. For simplicity, an optimistic 
complete overlap is assumed, and hence throughout this work, $f_{\rm sky, 
\times}$ denotes the smaller of the two sky coverages of the galaxy and HI 
redshift survey respectively. We use 15 $\ell$-bins between $\ell = 1$ and $\ell 
= 1000$, logarithmically spaced with $\Delta \log_{10} \ell = 0.2$.

\begin{table}
\begin{center}
    \begin{tabular}{ | c | c | c | c | c | c | c | p{5cm} |}
    \hline
 Astrophysical &  & Cosmological & & \\
 \hline 
 log ($v_{\rm c,0}$/km s$^{-1}$) & 1.56  & $h$ & 0.71  \\
 $\beta$ & -0.58 & $\Omega_m$ & 0.28 \\
 $Q$ &  4.6 & $\Omega_b$ & 0.0462 \\
 &  &  $\sigma_8$ & 0.81\\
  &  & $n_s$ & 0.963 \\
 \hline \\
    \end{tabular}
\end{center}
\caption{Fiducial values of the astrophysical and cosmological parameters 
considered. {{Astrophysical parameters come from the best-fitting values of the 
halo model for neutral hydrogen (Paper II), and that of the the galaxy 
$Q$ parameter from the `blue5' galaxy sample in \citet{cresswell2009}. The 
cosmological parameters are in good agreement with most available observations, 
including the latest Planck results \citep{planck}.}}}
 \label{table:astrocosmo}
\end{table}

We use a Fisher forecasting formalism to place constraints on the cosmological 
and astrophysical parameters, given the experimental configuration under 
consideration. The Fisher matrix is computed as follows:
\begin{equation}
F_{ij} = \sum_{\ell} \frac{1}{(\Delta {C_{\ell, \times}})^2}\frac{\partial 
C_{\ell, \times}}{\partial p_i} \frac{\partial C_{\ell, \times}}{\partial p_j}
\label{fisher}
\end{equation}
where the sum is over the range of $\ell$'s probed, and the $p_i$'s denote the 
individual parameters.

The following parameters are used for the computation of the cross-power 
spectrum of HI and galaxy surveys:
\begin{enumerate}
\item  The HI-based astrophysical parameters include $v_{c,0}$, $\alpha$, and 
$\beta$ used in the $M_{\rm HI} (M)$ relation, and the two parameters $c_{\rm 
HI, 0}$ and $\gamma$ for the HI profile,

\item the galaxy astrophysics contains the three parameters $b_0$, $b_{\rm 
ls}$, $Q$ and $A$ used in the large-scale and the scale-dependent part of the 
bias respectively, and

\item the cosmological parameters are the Hubble parameter $h$, the baryon 
density $\Omega_b$, the spectral index $n_s$, the power spectrum normalization 
parameter $\sigma_8$, and the cosmological matter density, $\Omega_m$.
\end{enumerate}

Of the HI astrophysical parameters, only two, viz.  the cutoff and the slope of 
the HI-halo mass relation, i.e. $v_{c,0}$ and $\beta$ are relevant for 
forecasting with HI intensity mapping surveys (see Paper I for details). While 
we use all the galaxy parameters to model the bias for various surveys, we vary 
only the parameter $Q$ encoding the scale-dependence of the bias.
Throughout the analysis, the cosmology adopted is flat, i.e. $\Omega_{\Lambda} 
= 1 - \Omega_m$. The fiducial values of the cosmological and astrophysical 
parameters are listed in Table \ref{table:astrocosmo}.

For calculating the standard deviations of the various cosmological and 
astrophysical parameters, we use a procedure similar to Paper I: we consider 
equal sized redshift bins of width $\Delta z \approx 0.5$ each, spanning the 
desired 
cross-correlation range in redshift, and evaluate the Fisher matrices $F_{ij}$ 
given by \eq{fisher} at the midpoints of each of the bins. The cumulative 
Fisher 
matrix for the $z-$range is derived from tomographic addition of the bins: 
$F_{ij, \rm cumul} = \sum_{\Delta z \in z} F_{ij}$, which is the sum of the individual Fisher matrices, $F_{ij}$ in each of the $z-$bins of width $\Delta z$ contained between 0 and $z$.
From the cumulative Fisher matrix, the standard 
errors in the parameters are computed for various cases. We ignore the 
effects of cross-correlations between individual bins and those between galaxies 
and HI in adjacent bins.

\section{Experiment combinations}
\label{sec:surveys}

For each galaxy survey, the specifications include the large-scale galaxy bias, 
$b_{\rm gal, ls}$, the selection function $\phi(z)$ and the total number density 
of galaxies, $n_{\rm gal}$. These as 
well as the survey properties of the HI surveys are listed together in Table 
\ref{table:surveyprops}.

 \begin{table*}
\begin{center}
    \begin{tabular}{ | c | c | c | c | c | c | c | c | c| p{1cm} |}
    \hline 
   &   Galaxy & &  & HI &  &   Cross-correlation & & \\
    Configuration & $n_g$ {(\rm amin$^{-2}$)} & $b_{\rm ls, gal}$ & $T_{\rm 
inst}$ (K) & $N_{\rm dish} $ & $D_{\rm dish} $ (m.)  & $f_{\rm sky, \times}$  &  
$z_{\rm 
bins}$\\ \hline
   CHIME-DESI  & 0.33  & 0.84/$D(z)$ & 50 & 1280 & 20  & 0.44 & [0.8, 1.2, 1.6] 
\\ 
     MeerKAT-DES  & 8 & $(1.07 - 0.35 z)^{-1}$ &  29 & 64 & 13.5  &  0.12 & 
[0.5, 1.0, 1.4] \\
     SKA I MID-LSST &  26 & $1.46(1 + 0.84 z)$ &  28 & 190 & 15  & 0.48 & 
[0.082, 0.58, 1., 1.5, 2., 2.3, 3.06] \\ \hline
    \end{tabular}
\end{center}
\caption{Various experimental configurations considered in this work.}
 \label{table:surveyprops}
\end{table*}

\subsection{CHIME and DESI}

\begin{figure}
\begin{center}
\includegraphics[scale = 0.6, width = \columnwidth]{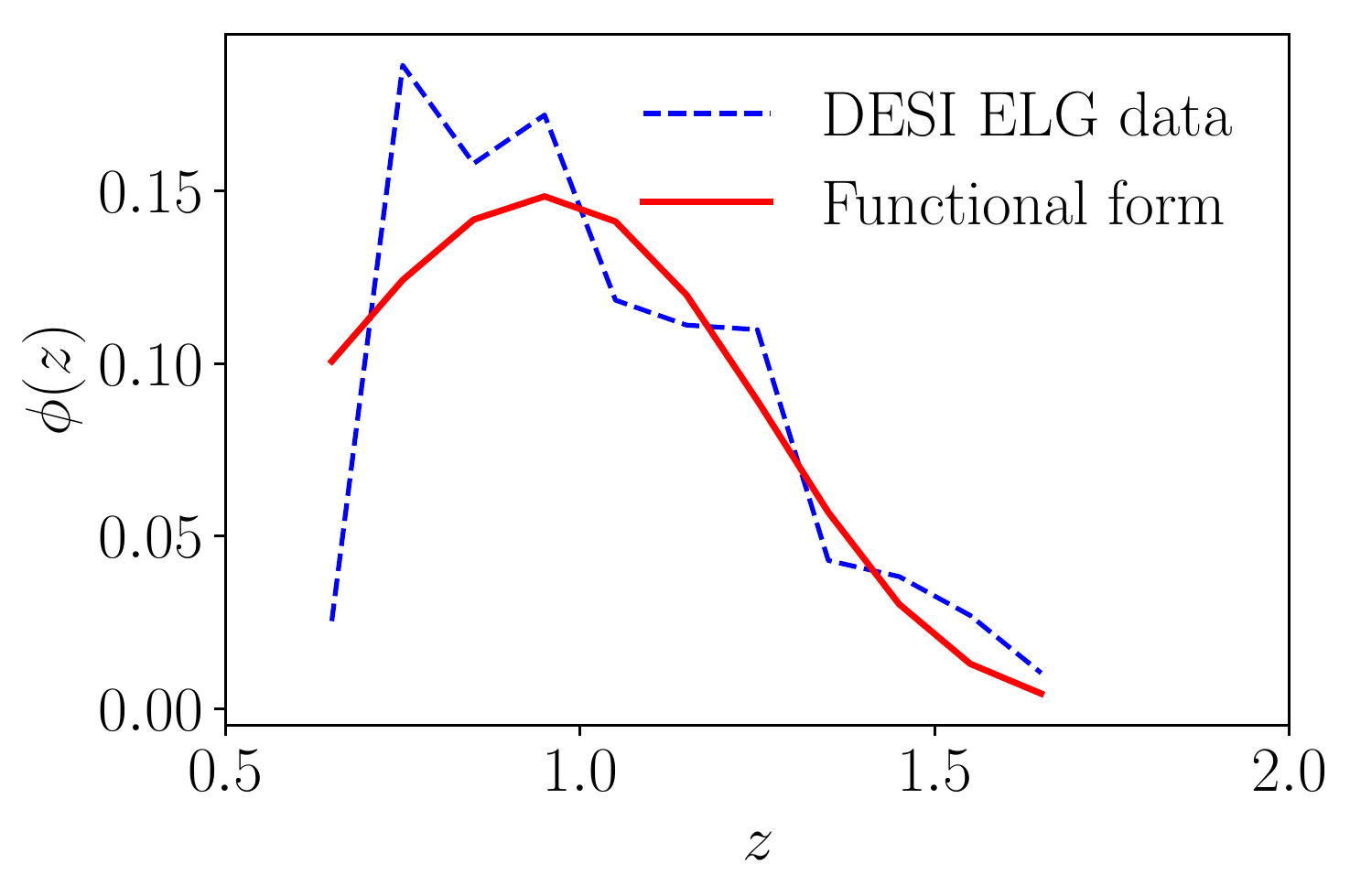}
\caption{Redshift selection function constructed from the forecasted number 
counts for DESI ELG galaxies in \citet{desi2016}, along with its fitted 
functional form represented by \eq{desisel}.}
\label{fig:selfunc}
\end{center}
\end{figure}

The redshift coverage of the cross-correlation is $0.8 < z < 1.8$. The CHIME 
autocorrelation survey runs over $z \sim 0.8 - 2.5$.
The DESI sample is assumed to correspond to the Emission Line Galaxy  (ELG) 
survey \footnote{This is the largest sample of galaxies for which DESI will 
obtain spectroscopic redshifts over $z \sim 0.6-1.8$.}, with the bias factor 
$b_{\rm gal, ls} = 0.84/D(z)$ where $D(z)$ is the 
growth factor.   The selection 
function for DESI is constructed by numerically fitting to the number counts in 
the ELG forecasts over $z \sim 0.6 - 1.8$, Table 2 of \citet{desi2016}. It is 
found that this selection function can be modelled as:
\begin{equation}
 \phi(z) \propto (z/z*)^2 \exp(- (z/z_0)^{\beta})\
 \label{desisel}
\end{equation} 
where $z* = 1.96, z_0 = 1.14$ and $\beta = 4.36$. This selection function, as 
well as the raw number counts forecasted for DESI ELG galaxies, are plotted in 
Fig. \ref{fig:selfunc}. The surface number density of 
galaxies is $n_{\rm gal} \approx 0.33$ arcmin$^{-2}$ (corresponding 
to roughly 1200 galaxies per square degree), which is consistent with the 
estimates for the numbers of ELG targets in \citet{desi2016}.

\begin{figure}
\begin{center}
\includegraphics[scale = 0.6, width = \columnwidth]{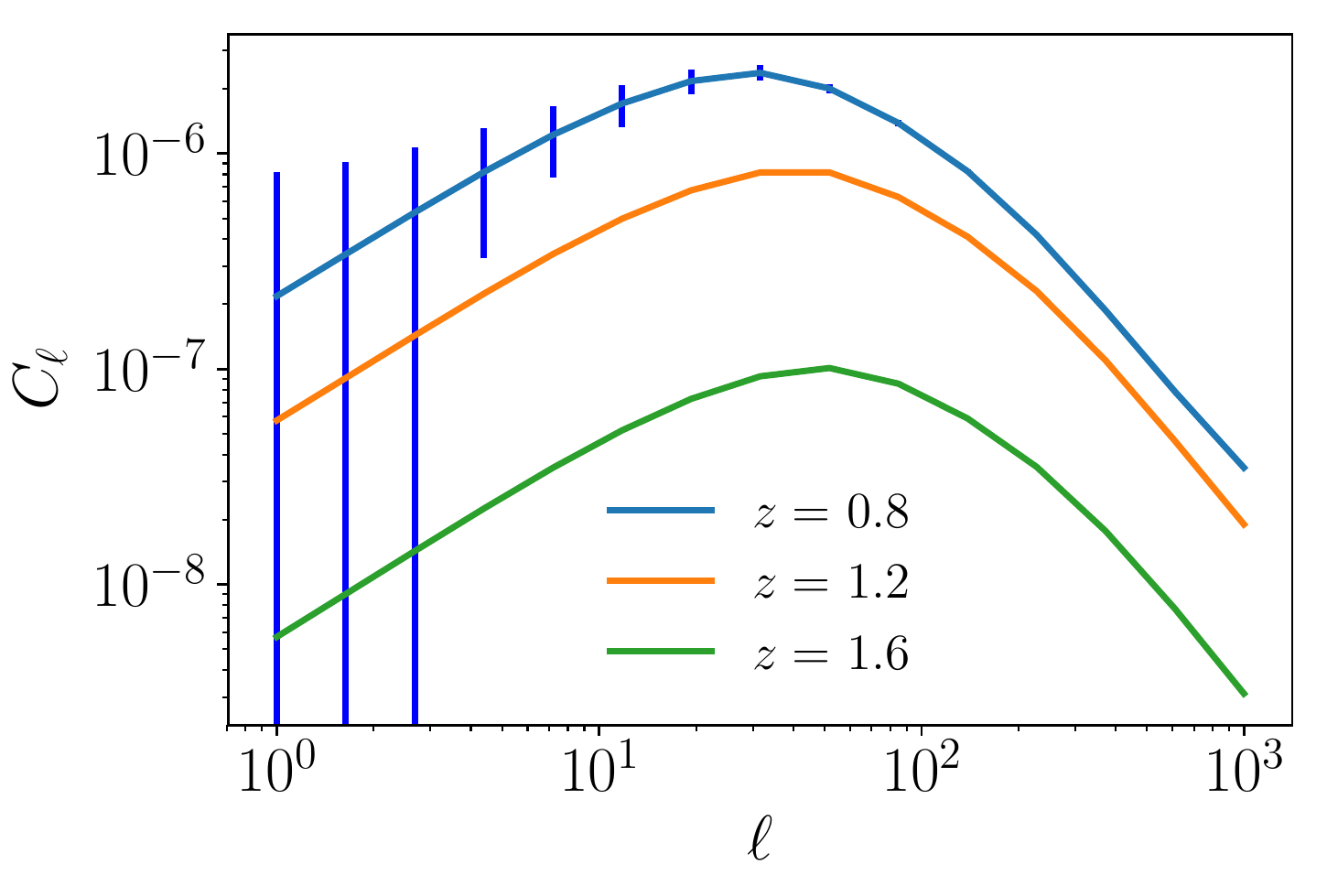}
\caption{Cross-correlation power spectra for a CHIME-DESI like survey, at the 
three 
redshifts of interest. Error bars indicate the expected standard deviation of 
the angular power spectrum (\eq{clerror}) at 
the lowest redshift ($z \sim 0.8$).}
\label{fig:clexample}
\end{center}
\end{figure}

The observing time $t_{\rm 
pix}$ is assumed to be 1 hour (per pixel). The sky coverages for the individual 
surveys are taken as $f_{\rm sky}$ (CHIME) =  0.61 (corresponding to $25000$ 
deg$^2$) 
and $f_{\rm sky}$ (DESI) =  0.44 (corresponding to $18000$ deg$^2$) and the 
DESI value is assumed for $f_{\rm sky, \times}$. We consider equal-sized 
redshift bins of width $\Delta z = 0.5$ each.

Plotted in Fig. \ref{fig:clexample} are the cross-correlation angular power 
spectra (computed following \eq{clcross}) for the 
CHIME-DESI like configuration at three mean redshifts 0.8, 1.2 and 1.6. At the 
lowest redshift, the error bars indicating the expected standard deviation on 
the power spectrum (from \eq{clerror}) are also
plotted.

\begin{figure*}
\includegraphics[scale = 0.6, width = 
\textwidth]{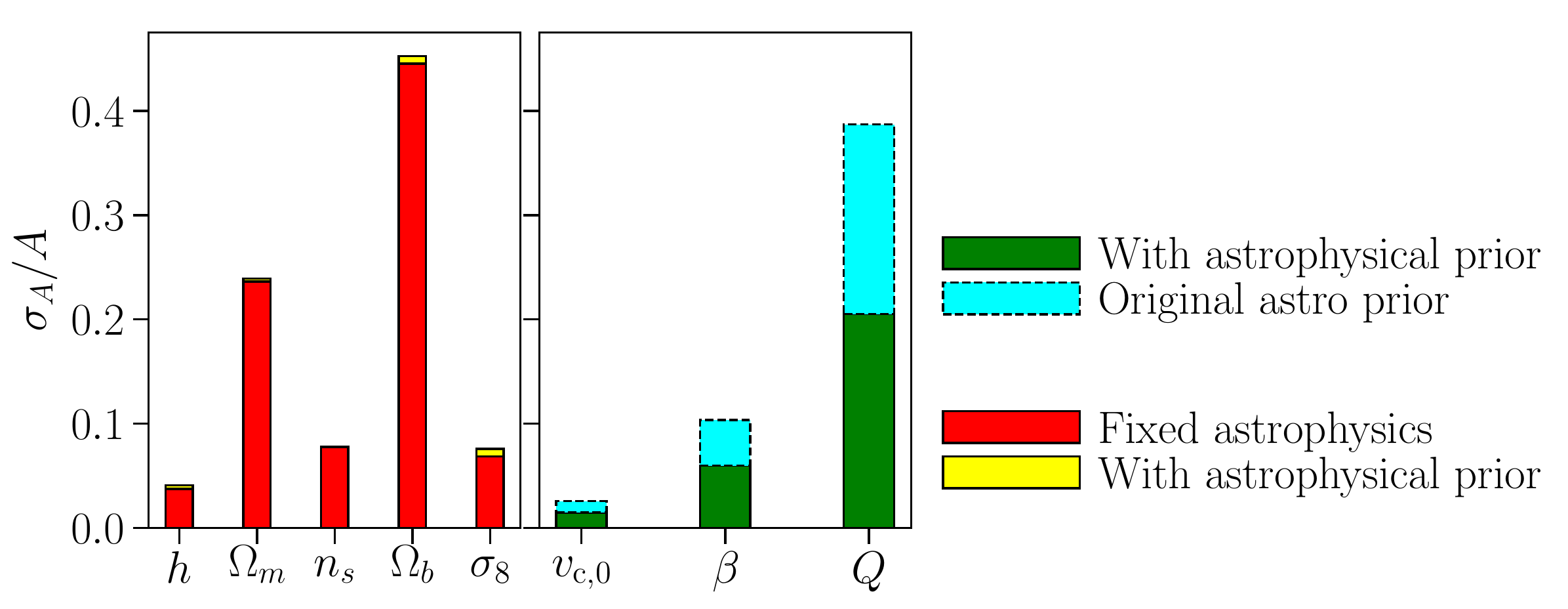}
\caption{Cross-correlation forecasts on astrophysics and cosmology for a 
CHIME-DESI like survey. Fractional errors, $\sigma_A/A$ are plotted with $A = 
\{h, \Omega_m, n_s, \Omega_b, \sigma_8, v_{c,0}, \beta, Q\}$, using information 
from all the redshift bins available over the combined dataset. The left panel 
shows the constraints on the cosmological parameters (i) without marginalizing 
over astrophysics and (ii) marginalizing over astrophysics but including a 
prior 
based on the current knowledge of the astrophysical parameters. The right panel 
plots the constraints on the astrophysical parameters for the case of marginalizing over cosmology but adding 
the astrophysical prior. The extent of the astrophysical prior assumed is 
plotted as the cyan band for each case in the right panel.}
\label{fig:forecasts}
\end{figure*}

In Fig. \ref{fig:forecasts} are plotted the cumulative fractional errors 
(combining all the redshifts under construction) on the 
forecasted cosmological and astrophysical (both HI and galaxy) parameters in 
the following cases: (a) with fixed cosmology, i.e. without marginalization 
over 
the cosmological parameters, (b) with fixed astrophysics, and (c) marginalizing 
over the galaxy and HI astrophysics, assuming a prior on the astrophysical 
parameters coming from the current knowledge of the HI and galaxy data.  The 
extent of these astrophysical priors are plotted in violet.  The HI 
parameters are assumed to have the best-fit standard deviation values 
constrained by the presently available data (see Table 3 of 
\citet{hparaa2016}).{The galaxy 
parameter $Q$ is taken to have a standard deviation of $1.78$, following the 
discussion for 
the `blue5' galaxy sample in \citet{cresswell2009}.}

The left panel  of Fig. \ref{fig:forecasts} shows that the constraints in the 
CHIME-DESI cross-correlation case improve on the 
 corresponding auto-correlation constraints using a CHIME-like 
configuration alone, by factors of about 1.1-2 depending on the cosmological 
parameter under consideration (comparing to Fig. 7 of Paper I). It is notable 
that this improvement occurs even though the redshift coverage of the 
cross-correlation is only about half that of the autocorrelation survey, and 
illustrates the extent to which adding the galaxy survey information helps 
improve
the cosmological constraints.

For all the three astrophysical parameters, there is a marked improvement on the current knowledge of the 
astrophysics from the cross-correlation information, as represented by the 
relative magnitudes of the violet and cyan/green bars. The  constraints improve by 
factors 
of 3.5, 3.1 and 3.3 respectively compared to their current priors.

\subsection{MeerKAT and DES}

\begin{figure*}
\includegraphics[scale = 0.6, width = 
\textwidth]{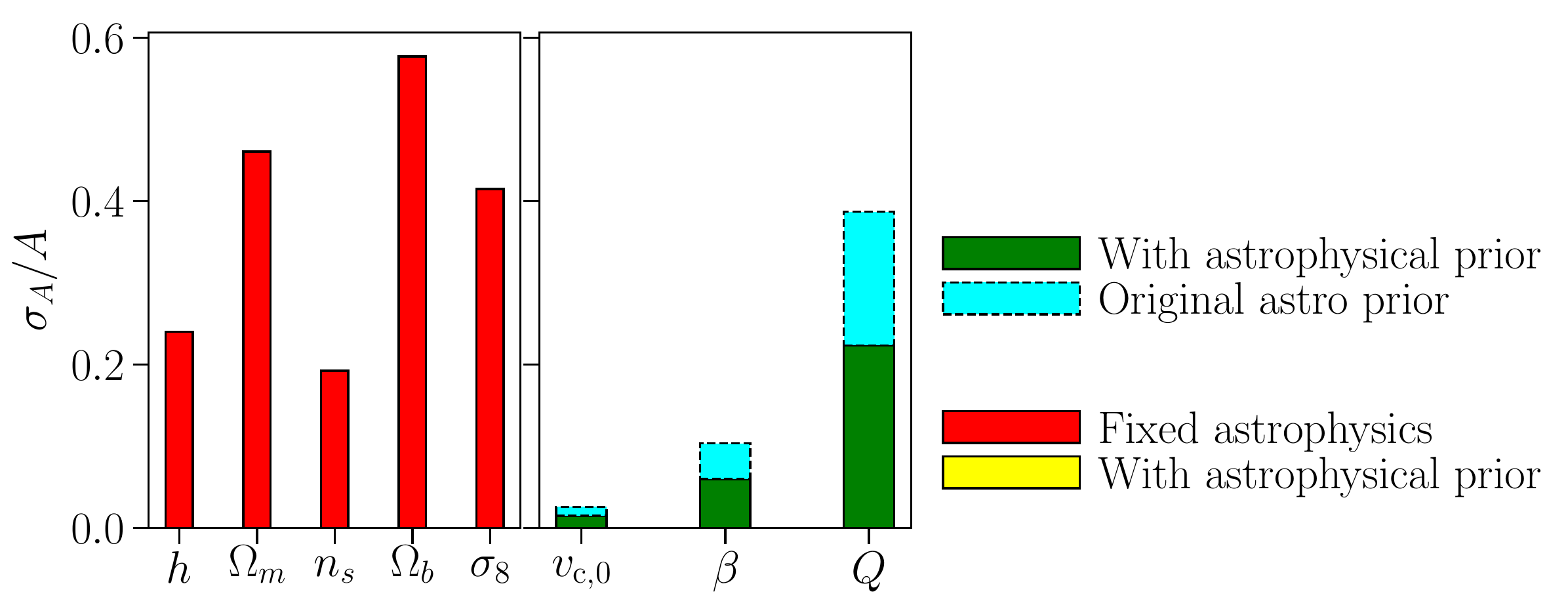}
\caption{Same as Fig. \ref{fig:forecasts}, for  a MeerKAT-DES like survey.}
\label{fig:meerkatdes}
\end{figure*}

\begin{figure*}
\includegraphics[scale = 0.6, width = 
\textwidth]{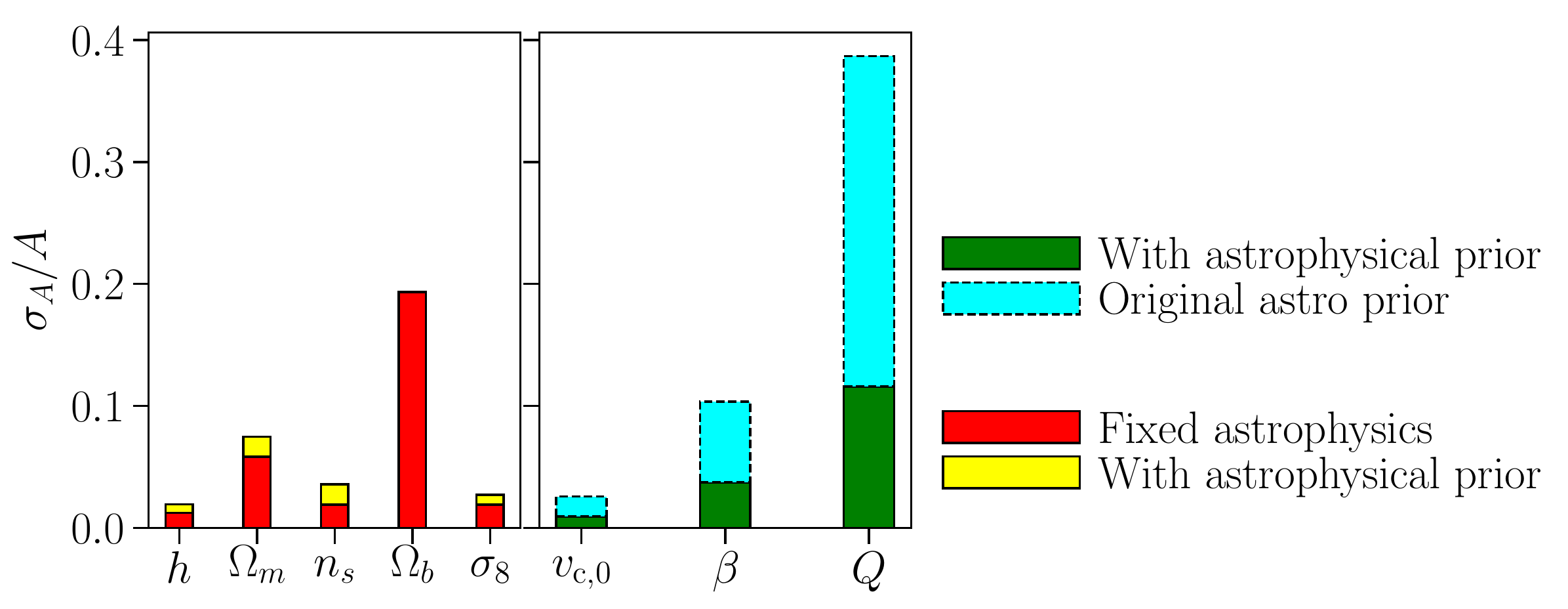}
\caption{Same as Fig. \ref{fig:forecasts}, for  an LSST-SKA like survey.}
\label{fig:lsstska}
\end{figure*}

 \citet{pourtsidou2016} discussed the potential for forecasting  lensing 
convergence 
parameters with a MeerKAT-DES survey in the southern hemisphere.\footnote{The 
MeerKLASS 
\citep[MeerKAT Large Area Synoptic Survey;][]{santos2017} proposes to 
investigate galaxy evolution and cosmology using a 4000 deg$^2$ overlap with 
the Dark Energy Survey (DES).} Here, we explore how such a cross-correlation 
survey could potentially constrain the cosmological and astrophysical (both 
galaxy bias and HI) parameters as in the previous case.

The redshift coverage of the survey is taken to broadly cover $z = 0.2$ to $z = 
1.4$. 
(Both DES and MeerKAT cover a similar redshift range, so the redshift overlap 
is stronger between these two surveys.) The sky coverage is assumed to be all 
of the DES survey, $5000 \  {\rm deg}^2$, thus assuming complete overlap. The 
galaxy bias for the DES galaxies is given by the fitting formula 
\citep{chang2016, pujol2016}: $b_{\rm gal, ls}^{-1} = 1.07 - 0.35z$. 
The redshift selection function for the DES galaxies is taken to have the form 
\citep[e.g.,][]{crocce2011}:
\begin{equation}
 \phi(z) \propto \left(\frac{z}{0.5}\right)^2 
\exp\left(-\left(\frac{z}{0.5}\right)^{1.5}\right)
\end{equation} 
The 
surface number density of the DES galaxies under consideration is $n_{\rm gal} 
= 
$ 8 arcmin$^{-2}$, consistent with the estimates in \citet{becker2016}. 

The fractional errors on the parameters considered are plotted in Fig. 
\ref{fig:meerkatdes}. Errors in the 'fixed astrophysics case' remain essentially 
unchanged from those with the astrophysical prior. 
The 
constraints on the cosmological parameters are of the same order of magnitude as those from the MeerKAT autocorrelation survey.  
The astrophysical constraints improve over the current knowledge 
of these parameters as 
seen by the relative magnitudes of the cyan and the green bars, by 
factors of about 1.5-2 for $v_{\rm c,0}, \beta$ and $Q$ respectively.

\subsection{SKA I MID and LSST}

The Large Synoptic Survey Telescope (LSST) survey parameters are taken to be 
\citep{abell2009}, see also \citet{chang2013,ferraro2018}: (i) the galaxy 
surface number density $n = 26$ arcmin$^{-2}$, and (ii) large scale bias 
$b_{\rm gal, ls} = 1.46(1+ 0.84z)$. The redshift coverage is from $z \sim 0-3$ 
which spans both the SKAI-MID (B1 and B2) bands. The redshift selection function 
of the survey is taken to be \citep{chang2013}:
\begin{equation}
\phi(z) \propto z^{1.28} \exp\left(-\frac{z}{0.41}\right)^{0.97}
\end{equation} 
The sky coverage of LSST is 
assumed to be 20000 deg$^2$ and that of the SKA-I is 25000 deg$^2$, and hence 
the LSST coverage is used for calculating $f_{\rm sky, \times}$ (assuming 
complete overlap).

This configuration leads to the tightest constraints on all the cosmological 
and astrophysical parameters as shown in Fig. \ref{fig:lsstska}, with all 
relative errors being about a few percent. It also leads to an substantial improvement in the 
astrophysical constraints as compared with those from the present data. 

The relative errors on the cosmological parameters reach values down to  $\sim 
0.01$ with 
this configuration. Constraints on $h$, $\sigma_8$ and $n_s$ with the 
astrophysical prior improve by factors 
of a few to ten, compared to the corresponding values from the SKA I - 
MID like 
autocorrelation survey alone (shown in Paper I), while those on 
$\Omega_m$ and $\Omega_b$ improve by factors 2-5. Further, the astrophysical 
parameters    
improve over their current priors by factors of 2.7, 2.8 and 3.3 for $v_{c,0}, 
\beta$ and $Q$ respectively.

\section{Conclusions} 
\label{sec:conclusions}
In this paper, we have explored combining upcoming HI intensity mapping surveys 
with wide field galaxy optical surveys to improve available constraints on 
astrophysical and cosmological parameters over $z \sim 0-3$ in the 
post-reionization universe. Using the $\Lambda$CDM cosmological 
parametrization, 
a halo model framework for HI driven by currently available data, and available 
optical 
galaxy parametrizations, we have studied the extent to which these 
constraints improve over their current uncertainties due to cross-correlation 
measurements. We also note the improvement in the constraints compared to those 
from the corresponding HI auto-correlation surveys alone.

For all three survey cases considered (a CHIME-DESI-like survey in the northern 
hemisphere, and (ii) a MeerKAT-DES-like and LSST-SKA-like survey in the 
southern 
hemisphere), we find that the cross-correlation leads to 
improvements in measurement of both astrophysical and cosmological parameters, 
though the extent of improvement depends on the parameter under consideration. The significant benefit of cross-correlation (particularly in the MeerKAT-DES-like and CHIME-DESI-like configurations) lies in the improvement of astrophysical constraints.
The halo model framework allows us 
to place realistic priors on the HI astrophysics from the currently available 
data.  
With the LSST-SKA combination, all the parameter constraints (both 
astrophysical 
and cosmological) reach levels below about 20 percent, even 
without the assumption of cosmological priors. 

The astrophysical forecasts for the HI and galaxy parameters improve 
substantially (by factors of a few) over their current priors, with 
the help of the cross-correlation measurements. This holds even in the presence 
of the additional galaxy 
parameter $Q$, which is seen to have comparable constraints though its prior 
knowledge is assumed to be more uncertain.
The cosmological forecasts improve by factors of about a few (depending on the configuration) over those from 
the corresponding autocorrelation surveys alone.

We note that the foregrounds, which may be the limiting systematic, are 
excluded from the noise calculation in the present study. However, in the case 
of cross-correlation measurements, the foregrounds for the two individual probes 
are expected to be significantly
uncorrelated and thus lead to negligible effects \citep[as shown for the case 
of 
non-smooth foregrounds in 21 cm cross correlations with LBG surveys in, 
e.g][]{navarro2015}. Recent studies \citep{foreground12019, foreground22019, breysse2019} 
describe ways in which the signal, in the presence of foregrounds, may be 
reconstructed from 21 cm intensity mapping data cross-correlated with other 
tracers, such as the CMB or optical (both spectroscopic and photometric) galaxy 
surveys. { A couple of caveats in this respect, however, are worth mentioning.

(i) First, we know that 21 cm experiments in autocorrelation suffer from bright foregrounds, which must be removed effectively, leading to a loss of low-$k$ modes (this effect is particularly prominent for the low-$k$ modes in the radial direction, which correspond to the largest scales). Hence, these Fourier modes are likely to be lost from the survey information, even if systematic biases caused by foregrounds can be disregarded. 

(ii) We have not explicitly modelled the effects of the galaxy photo-$z$ errors in the present analysis. Since fairly broad redshift bins are used (with $\Delta z \sim 0.5$), the effect of these errors may be largely mitigated for the case of the galaxy-galaxy autocorrelation. However, if foreground filtering results in the removal of radial modes in the 21 cm surveys up to some maximum (as described in the previous point), then there may be little overlap between the 21 cm radial modes and the galaxy radial modes that remain for the cross-correlation. This effect may need careful treatment when foreground cleaning or avoidance is being considered.
}

Just as in the previous study with auto-correlation data alone (Paper II),  we 
see 
that the overlap in redshift coverage is extremely 
important in tightening forecasts (due to more information coming from the 
addition of independent tomographic bins). Also, choosing similar sky area 
overlap between the HI 
and galaxy surveys (presently assumed to have complete overlap)  would lead to 
better constraints on the parameters. 
Extending these approaches towards intensity mapping with other emission lines 
(CO, CII) would enable 
us to potentially form a comprehensive picture of galaxy evolution at the 
scales 
of the 
ISM. Ultimately, combining both auto-and cross-correlation forecasts, possibly 
with cosmological priors from present and future CMB experiments, would provide 
the tightest possible constraints exploiting the synergy of CMB, HI and galaxy 
surveys. This would be a powerful tool to explore more parameters in 
cosmological models such as e.g., testing modifications to general relativity 
at 
the largest scales \citep[e.g.,][]{hall2013}  with future wide-field surveys.

\section{Acknowledgements}
We thank Stefano Camera, Ue-Li Pen and Ren\'{e}e Hlo{\v z}ek for useful 
discussions, {{and the anonymous referee for a detailed and helpful report that improved the content and quality of the presentation.}} HP's research was supported by the Tomalla Foundation. AA is supported by the Royal Society Wolfson Fellowship.

\bibliographystyle{mnras}
\bibliography{mybib}

\end{document}